# Innovating with Generative AI: A Human Bottleneck Framework

Julian De Freitas, Ayelet Israeli, Gideon Nave, Artem Timoshenko, Olivier Toubia

## Abstract

We propose a human bottleneck perspective for understanding how generative AI transforms the innovation process. The central premise is that many constraints traditionally plaguing the innovation process are cognitive and social in origin, rooted in how people generate ideas, evaluate novelty, and communicate through social systems. Generative AI does not act uniformly on these constraints. At each stage, it can deepen some bottlenecks while alleviating others, and predicting these outcomes requires understanding the underlying mechanisms of the constraint itself. We identify bottlenecks in four stages of the innovation process: ideation, screening and testing, preference measurement and consumer insight, diffusion, and market learning. By grounding analysis in human behavior rather than rapidly changing AI capabilities, we offer a framework for assessing whether new developments alleviate or intensify the bottlenecks that matter most at each stage. We also distinguish bottlenecks likely to narrow as capabilities improve from those rooted in enduring human constraints. We further discuss AI-related issues that cut across the entire innovation pipeline, challenging the very existence and structure of the traditional innovation process.

**Keywords**: generative AI, innovation, new product development, digital twins, human-AI complementarity

## 1. Introduction

Generative AI, or GenAI, has been incorporated at every stage of the innovation process. Firms use large language models (LLMs) to mine social listening data for consumer insights, generate product concepts at scale, simulate market reactions to new features, and produce launch communications in minutes. An empirical literature is forming around each of these applications, spanning idea generation (De Freitas et al., 2025), consumer simulation (Arora et al. 2025; Toubia et al., 2025), preference measurement (Brand et al., 2026), aesthetic design (Wu et al., 2026), customer need extraction (Timoshenko et al., 2026), and diffusion (Park et al., 2023). But empirical findings are accumulating faster than organizing frameworks for interpreting them. What is missing is a principled basis for asking which developments are likely to matter, which human constraints they are operating against, and what the division of labor between generative AI and human capability should be. As we will argue, the answer to all three questions still depends, in large part, on features of human involvement in the innovation process that predate generative AI.

Recent work has begun to organize this terrain. De Freitas et al. (2025) present a psychologically-inspired account of how LLMs can aid in ideation, the first stage of the innovation process. Cillo and Rubera (2025) propose a roadmap structured around stages of the entire innovation process (developing, testing, communicating, engaging), cataloging research questions at each stage. The current work adds a similarly broad framework with a different organizing logic. Rather than structuring the analysis around stages or AI capabilities, which both shift as the technology evolves, we ground it in stable "human bottlenecks" that cut across the innovation process (Section 2).

Each subsection follows the same structure. We outline the main challenge at a given stage of innovation, identify its human bottlenecks, and examine how generative AI affects them, including instances where AI exacerbates the challenges. We then discuss how to appropriately intervene on the bottleneck. A key practical implication of the framework is that the right intervention depends on which mechanism is operative. Some bottlenecks can be fully alleviated with  generative AI at the individual, model, or process level, while others are rooted in enduring human psychological, social, and organizational constraints, and they are unlikely to yield to any foreseeable technological advance. See Table 1 for an overview.

Next, in Section 3, we zoom out to reflect on systemic transformation at the level of the entire innovation process. We identify and discuss issues related to the use of GenAI in innovation that cut across the entire process. We also consider the innovation process as we know it today and how it may evolve or be challenged due to GenAI. For example, we explore what innovating will look like in a world where GenAI systems are not just developers but also become users of innovations.

**Table 1.** The human bottleneck framework integrating generative AI into the innovation process.

| **Innovation stage** | **Human bottleneck** | **Influence of unchecked generative AI** | **Intervention** |
|---|---|---|---|
| **Ideation** | **Cognitive fixation.** Individuals anchor to familiar mental models, limiting access to novel ideas | Deepens fixation: LLMs cluster around typical outputs, and exposure narrows human retrieval further | Chain-of-thought prompting; lead-user programs and ethnography for ideas requiring lived experience |
| | **Design fixation**. Creators converge on surface features of exemplars, reducing originality across teams | Mode collapse: independent AI sessions draw from the same training distribution, homogenizing outputs industry-wide | Persona modifiers to push the model into varied distribution regions; AI-generated variants of exemplars that preserve core design elements while altering surface features |
| | **Psychological safety**. Self-censorship and status hierarchies suppress novel ideas in group settings | Reduces social judgment risk at generation, but status dynamics re-enter at evaluation; collaborative vs. competitive mode depends entirely on prompting | AI-mediated hybrid nominal process: private generation → anonymized pool → structured group review; AI enforces divergent (yes-and) and convergent (critical) phases separately |
| **Screening & testing** | **Novelty-averse evaluation**. Implicit bias against original ideas due to uncertainty aversion and prototype anchoring | Fluent AI outputs trigger false positives via cognitive fluency, inflating scores regardless of idea quality | Strip ideas to a common format before scoring; blind evaluators to authorship; sequence feasibility/novelty assessments deliberately to manage the mean-variance tradeoff |
| | **Cognitive load**. High idea volume pushes evaluators toward heuristics that disfavor novelty | Dramatically increases idea volume, amplifying reliance on novelty-averse heuristics | Combine LLM ratings with historical human expert ratings to triage submissions; surface only those requiring careful review |
| | **Shared mental model lock-in**. Panels develop implicit, self-reinforcing criteria for "promising" ideas | AI trained on historical decisions encodes and scales panel biases; rationale-accompanied recommendations reduce independent evaluator verification | Train models on post-launch outcomes rather than screening decisions; use AI to generate the strongest case for an idea before scoring; withhold AI rationales in screening contexts |
| **Preference measurement & consumer insight** | **Authenticity gap**. Preferences are constructed in context; non-normative behaviors govern adoption | Heterogeneity loss, prompt-induced confounding, and failure to reproduce irrational adoption behaviors (e.g. sunk cost fallacy, loss aversion) | Augment (not replace) human surveys with calibrated LLM data; unblind experimental design to the model; reserve non-normative adoption research for human methods and ethnography |
| | **Articulability gap**. Consumers cannot articulate preferences for products they have never encountered | AI can only surface preferences expressed in language; AI-moderated interviews may systematically | Ethnographic/observational methods remain the only route to "hidden gems"; use AI-generated stimuli to surface tacit manager intuitions; use fine- |
| | | narrow the vocabulary of elicited preferences | tuned models for scalable need-statement extraction |
| **Diffusion & market learning** | **Social diffusion modeling**. Adoption is a social process difficult to model across multiple segments | Inherits and amplifies individual simulation errors as they propagate through interaction effects | No current AI fix; AI screens plausible adoption paths; consequential calls remain human, supported by lead-user and ethnographic work |
| | **Attention & aggregation**. Post-launch signal volume exceeds human capacity; selection bias obscures key signals | Substantially alleviates volume constraint but shifts bottleneck to prioritization | Scalable LLM-based need extraction from unstructured text; architect triage to surface outliers and weak signals, not just high-frequency themes |
| | **Motivated interpretation**. Organizations interpret market signals to confirm prior decisions | AI surfaces negative patterns neutrally, but fluent summaries are easier to selectively cite in support of prior conclusions | No empirically validated intervention yet; open question whether AI synthesis amplifies or counteracts confirmation bias |

## 2. Generative AI and Innovation Bottlenecks

The innovation literature is vast, and any single framework can only engage selectively with it. The bottlenecks we identify below are not exhaustive. Instead, we chose major challenges that illuminate a recurring pattern: whether generative AI alleviates, reshapes, or intensifies a given innovation constraint depends on the human and organizational processes that created that constraint in the first place.. We treat these examples as representative of a broader logic rather than a complete inventory. We structure our discussion around the traditional stages of the innovation process: ideation, screening and testing, consumer insight, and diffusion and market learning (Hauser et al., 2006).

## 2.1. The Ideation Stage

Ideation, the process of generating new concepts to satisfy a specific goal, sits at the heart of marketing research and practice. The fundamental challenge is producing ideas that are simultaneously original and appropriate. They must provide a meaningful deviation from what already exists while remaining practical to actually solve the problem at hand. These two properties often pull in opposite directions: an idea can be striking yet not useful or feasible, or highly valuable yet derivative of past work. Compounding this difficulty, the aim of ideation is not to accumulate a large pool of reasonably good ideas but to surface the very best ones. A single exceptional idea is worth more than dozens of average ones (Girotra et al., 2010). Consequently, ideation is fundamentally a search for outliers rather than a pursuit of consistency.

We highlight three human bottlenecks in ideation. Two are forms of fixedness operating at different levels: *cognitive fixation* at the individual level and *design fixation* at the social level. The third concerns the group dynamics that govern ideation in organization settings, principally *psychological safety* and the related choice between collaborative and competitive norms.

### 2.1.1. Bottleneck #1: Cognitive Fixation

At the individual level, *cognitive fixation* occurs when existing mental models–built from experience, knowledge, and habit–make some ideas accessible while rendering others effectively elusive. This manifests in two common ways: struggling to see novel uses for familiar concepts (known as "functional fixedness"), and falling into "mental sets" where one solution path progressively narrows the search space (Duncker, 1945; Luchins, 1942). Expert myopia is a related manifestation of this constraint. While informed outsiders might view a problem with fresh eyes, deep domain experts often fail to generate novel ideas that deviate from their expertise (Wiley, 1998). This is not for a lack of effort, but because their highly entrenched mental models efficiently and prematurely route their attention toward familiar solution spaces only.

The literature on lead users provides another powerful illustration of how deeply rooted these fixations are. Some of the most novel product ideas tend to come from users whose needs are so extreme that existing solutions have already failed them, forcing them to develop alternatives from scratch (Von Hippel, 1986, 1988; Holmes, 2018). This acute, lived experience is one of the few natural ways humans achieve true mental model escape. Such knowledge is built up through hands-on practice with the unsolved problem rather than through language (aka tacit knowledge; Polanyi, 1966), likely making it resistant to substitution by AI systems trained on text. The rarity of this phenomenon also highlights just how difficult it is for standard ideators to bypass cognitive and design fixation.

#### *How Generative AI Interacts with the Bottleneck*

When used naively (that is, by passively adopting model outputs with minimal engagement; Shaw & Nave, 2026), generative AI can deepen *cognitive fixation* at the individual level. LLMs are trained to generate text sequences by predicting the next token given the preceding context, so early outputs in an ideation session anchor the model toward related concepts, analogous to how activated memories anchor human ideation. Post-training alignment via reinforcement learning from human feedback can compound this: if annotators systematically favor familiar, conventional responses, the models are trained toward a restricted set of typical outputs. This tendency to provide what is typical is not confined to the model. When a human ideator reads the model's outputs, those outputs may dominate the retrieval process in human memory, suppressing access to alternatives the ideator could otherwise have generated (Smith et al., 1993). Naive use therefore deepens fixation through a two-step mechanism–the model concentrates its outputs around typical ideas, and exposure to those outputs then narrows the human's own retrieval.

### *Intervention*

For *cognitive fixation* within individuals, the target of intervention can be the model itself, and structured prompting can address the specific architectural mechanism. Chain-of-thought prompting asks the model to first generate short idea summaries, then explicitly revise them to be bolder and more distinct before expanding. This prevents early elaboration from constraining subsequent output and counteracts the typicality bias from alignment training. Notably, this intervention works for LLMs but not for humans: chain-of-thought prompting substantially increases within-session idea diversity in these models (up to a point, see Meincke et al., 2024a) but has essentially no effect on the diversity of ideas from human ideators (Deng et al., 2026). Once a line of thinking is established in a human mind, explicit instructions to diversify are largely ineffective. This pattern is consistent with ironic process theory, where consciously trying to avoid a mental representation keeps it active (Wegner et al., 1987), as well as with the part-list cuing phenomenon (Smith et al., 1993). LLMs, optimized through post-training to follow meta-instructions, do not share this limitation. Chain-of-thought prompting is therefore an intervention on the generative AI system, not on human cognition, and should be understood as such.

Finally, notice that the above intervention operates within the space of ideas that existing knowledge already makes accessible, while none surfaces ideas that require lived experience of an unsolved problem. Insights from lived experience remain the province of lead users, and the fix here is organizational rather than technological—deliberate investment in lead-user programs and ethnographic methods that put researchers close enough to unsolved problems for genuine mental-model escape to occur.

### 2.1.2. Bottleneck #2: Design Fixation

Fixedness also plays out at the social level as creators learn from one another. This matters because learning from successful exemplars is a fundamental driver of quality improvement in creative work. Exposure to high-performing solutions helps ideators discern effective strategies, align with evolving standards, and build on proven concepts (Von Hippel, 1986). But this same exposure triggers *design fixation*: exemplar exposure makes specific surface features highly accessible in memory, disproportionately shaping what designers retrieve during ideation. In aesthetic design, the result is convergence toward the look of the exact exemplars rather than integrating the core elements that make these designs work (Jansson & Smith, 1991; Smith et al., 1993). For example, in crowdsourcing contexts, design fixation creates a well-documented quality-originality tradeoff: open contests, where designers can view leading submissions, yield higher average quality but substantially lower originality; blind contests restore originality but yield lower quality than open contests by cutting off exposure to high-performing exemplars (Luo & Toubia, 2015; Hofstetter et al., 2021; Koh & Cheung, 2022; Wu et al., 2026).

### *How Generative AI Interacts with the Bottleneck*

The way generative AI interacts with design fixation takes a different form from cognition fixation but follows the same logic. Recall that in human creative work, design fixation arises when exposure to specific exemplars causes convergence on their surface features. LLMs exhibit the model-scale analog of this phenomenon, in which the implicit "exemplar set" is the centralized training distribution itself rather than any observed competitor work: pre-training captures the aggregate distribution of human knowledge rather than the distinct distributions that characterize individual minds, and then post-training alignment further compresses toward central tendencies. The result is that independent generative AI sessions sample from the same centralized distribution, producing outputs that cluster regardless of superficial variations in the prompt. This mode collapse has been documented across dozens of open and closed-source models (Jiang et al., 2025; Deng et al., 2026). Its practical consequence is the tragedy-of-the-commons dynamic: widespread generative AI adoption, even when generative AI output is solely used as an exemplar by human designers, narrows the collective solution space explored across an industry even as individual productivity rises (Doshi & Hauser, 2024; Meincke et al., 2024b,c; Hao et al., 2026).

### *Intervention*

Design fixation calls for a different intervention than cognitive fixation, given the different operative mechanism. *Design fixation* operates across individuals, and the model-scale analog of an "exemplar set" is the centralized training distribution itself. One promising intervention here is "persona modifiers", an intervention that prompts the model to respond as if from the perspective of a specific person (e.g., "a retired politician who has campaigned for library funding" or "a software engineer who collects stamps"), although not all persona modifiers are equally effective (Deng et al., 2026). This pushes the model into different regions of its training distribution, and effectively counters social homogeneity of ideas across sessions by injecting varied cues. This approach can also help address expert myopia, because the model can be asked how an uninformed outsider would approach the problem, forcing the model past the entrenched biases of the field.

Because *design fixation* occurs between humans, in which designers converge on surface features (e.g., how to render designs visually) rather than elements that are core to the design (e.g., motifs, composition), other relevant interventions are on the creative system through which information flows between humans. AI intermediation can separate surface features of designs from their more core elements by showing designers AI-generated variants of successful exemplars instead of the originals (Wu et al., 2026). Each variant retains the core elements of the original design while tweaking the visual execution, so that designers can absorb the core aspects of what makes a design work rather than simply copying its look.

### 2.1.3. Bottleneck #3: Psychological Safety

A third class of bottlenecks operates at the level of group dynamics in organizational ideation settings. Even when individual ideators have escaped cognitive fixation and are not anchored to specific exemplars, the social context of ideation introduces its own constraints. The main one is *psychological safety*: in organizational settings, ideators frequently self-censor not because a novel idea is cognitively inaccessible but because they anticipate social judgment (Diehl & Stroebe, 1987). The optimization target shifts implicitly from novelty to acceptability. Junior employees can be especially susceptible when seniors are present, and status and hierarchy can influence both who speaks and whose ideas are taken seriously, independent of idea quality (Keum & See, 2017). A second, related constraint concerns the structure of group interaction itself. Research on collaborative versus competitive ideation mindsets shows that whether participants are set up to critique or build on each other's ideas produces systematically different outcomes: "yes-and" norms generate greater associative breadth, while critical norms drive incremental refinement of a narrower set of candidates (Nijstad & Stroebe, 2006). Neither mode is uniformly superior, but the default structure of most organizational ideation sessions is rarely chosen deliberately with this tradeoff in mind.

### *How Generative AI Interacts with the Bottleneck*

Generative AI interacts with the psychological safety bottleneck asymmetrically. On the psychological safety dimension, externalizing ideation to an AI system removes the immediate risk of social judgment: people will articulate ideas to a model that they would not voice in a meeting (Wang & Chen, 2022). This is a genuine advantage and one of the more reliable social contributions of AI-assisted ideation in organizational contexts. However, the benefit is easily displaced rather than eliminated: if the AI-generated ideas are then presented and evaluated in a hierarchical group setting, status dynamics re-enter at the selection stage. The bottleneck has moved from generation to evaluation, without being resolved. On the collaborative-versus-competitive dimension, the model will build on ideas or critique them depending on how it is prompted, probably with an intrinsic sycophantic tendency to go in the user’s direction (Ranaldi & Pucci, 2023). The choice of ideation mode remains a deliberate design decision for the organization, not a property of the tool.

### *Intervention*

Group-dynamics bottlenecks require interventions in meeting architecture rather than model prompting alone. A useful structure is an AI-mediated hybrid nominal process: participants first use the model privately to generate and elaborate on ideas, then submit outputs to an anonymized pool where the model removes source cues, standardizes presentation, clusters near-duplicates, and preserves unusual features. Group discussion begins only after this pool exists. This separates idea generation from social display, reducing production blocking (resulting from the constraint that only one person can speak at a time in a group session - Diehl & Stroebe, 1987), evaluation apprehension, and status-based convergence while preserving later opportunities for recombination. The design builds on evidence that individual-then-group structures can outperform fully interactive brainstorming, and that electronic brainstorming systems can improve ideation by enabling parallel and relatively anonymous contribution (Connolly et al., 1990; Gallupe et al., 1992; Girotra et al., 2010). It is also consistent with the psychological-safety logic that people contribute more fully when interpersonal risk is reduced (Edmondson, 1999).

The same hybrid protocol can also separate divergent and convergent modes. During divergence, the AI model can act as a “yes-and” facilitator, prompting participants to build on anonymized ideas, inviting input from quieter or lower-status members, and discouraging premature evaluation. During convergence, it can switch into a critical-refinement role by surfacing assumptions, failure modes, and feasibility-preserving variants. This sequencing matters because psychological safety and critical evaluation are complements, not substitutes; they belong at different points in the process. GenAI can help enforce that boundary, but it cannot create psychological safety by itself. If senior actors dominate after ideas are revealed, the organizational bottleneck has simply moved downstream.

## 2.2. The Screening and Testing Stage

Screening and testing is the process of deciding which ideas to advance and which to abandon. Evaluators must judge ideas under uncertainty: before the market has spoken, before the product is built, and often before the idea itself is fully specified. In many organizations, this judgment is made not by individuals but by panels (stage-gate committees, portfolio reviews, and investment boards)--a setup widely treated as the gold standard for evaluating early-stage ideas. Screening is therefore an active force shaping which innovations reach the market, and its failures are, by construction, invisible.

Screening is subject to at least three kinds of bottlenecks. Two operate at the individual level — *novelty-averse evaluation*, which systematically discounts original ideas, and *cognitive load*, which compounds reliance on heuristic shortcuts. The third, shared *mental model lock-in*, operates at the group level.

### 2.2.1. Bottleneck #1: Novelty-Averse Evaluation

Even when people explicitly endorse creative thinking, they implicitly associate novelty with impracticality and uncertainty, triggering heuristic responses that disfavor genuinely original ideas at the screening stage–a bottleneck we term novelty-averse evaluation (Mueller, Melwani, & Goncalo, 2012). Loss aversion applies its own pressure, making the downside risk of a novel idea feel more salient than its potential upside (Lane et al., 2022). These biases reflect fundamental cognitive mechanisms that operate automatically in anyone making judgments under uncertainty. They persist in stage-gate scorecards, portfolio review panels, and the aesthetic judgments of experienced managers because they are cognitively cheap heuristics that work well in stable, incremental contexts yet fail systematically in novel ones (Potts 2010).

#### *How Generative AI Interacts with the Bottleneck*

Generative AI exacerbates *novelty-averse evaluation* through its cognitive-fluency channel. Because LLM-generated ideas are fluent and well-structured, their polish can be mistaken for quality (Lee & Chung, 2024), triggering  false positives in human evaluators. Evaluators relying on fluency as a proxy for merit may favor AI-produced ideas not because they are better but because they are easier to process. The human bottleneck is unchanged; what changes is the supply of stimuli that trigger it.

*Intervention*

For *novelty-averse evaluation*, the appropriate target is the structure of the evaluation task. We suggest two interventions. First, because the AI-specific exacerbation operates through the fluency channel (LLM outputs are optimized for the very surface qualities evaluators use as proxies for merit), interventions must decouple presentation from evaluation. Stripping ideas to a common format before scoring, blinding evaluators to authorship (human versus AI), or having a model rewrite all submissions into a uniform style before review removes the differential fluency advantage that AI-generated stimuli would otherwise enjoy. The bottleneck is not eliminated, but the supply of stimuli susceptible to triggering it is neutralized. Second, Grumbach et al. (2026) show that the order in which evaluators surface feasibility and novelty assessments produces a mean-variance tradeoff in selection: assessing feasibility before novelty yields higher mean innovation but lower variance, while assessing novelty before feasibility yields the reverse. AI augmentation can shape these sequencing heuristics, so the design of the human-AI interface becomes a deliberate organizational decision rather than a default to be inherited from existing stage-gate practice.

### 2.2.2. Bottleneck #2: Cognitive Load

Screening compounds the above biases by imposing capacity constraints. Screeners must typically choose a few winners from a large volume of candidates, and assessing highly novel ideas requires more cognitive resources than familiar ones because novel ideas are more difficult to interpret and categorize (Rindova & Petkova, 2007). Cognitive load shifts evaluators from systematic to heuristic processing, increasing reliance on cues like fluency, source expertise, and surface coherence, holding the quality of the idea constant (Petty & Cacioppo, 1986; Evans, 2003; Reber, Schwarz, & Winkielman, 2004). In practice, idea evaluators are less likely to prefer highly novel ideas under high load (Criscuolo et al., 2017). In sum, cognitive load does not have its own directional bias against novelty, but lowers the threshold at which the novelty-averse heuristics take over, making it a capacity multiplier on the first bottleneck rather than a separate source of bias.

### *How Generative AI Interacts with the Bottleneck*

Generative AI exacerbates the *cognitive-load* bottleneck directly, by dramatically increasing the number of ideas flowing into the evaluation pipeline. Because cognitive load is already a capacity multiplier on novelty aversion, this volumetric pressure also indirectly worsens the first bottleneck — evaluators under higher load lean more heavily on the heuristics that disfavor novelty.

### *Intervention*

For the *cognitive load* bottleneck, the appropriate intervention is on the architecture of the screening pipeline. Kireyev et al. (2025) demonstrate one such architecture using data from 153 ideation contests and over 74,000 submissions. By combining LLM-generated ratings with historical human expert ratings in a prediction model, the ideation platform could identify all sponsor finalist choices by reviewing 28.4% fewer submissions relative to sorting by the average of expert scores. Most of this gain comes from re-weighting historical human ratings to align with sponsor preferences, rather than the LLM signal. The intervention does not reduce the absolute volume of submissions, but it optimally selects the items that need careful review, lowering effective cognitive load without delegating substantive judgment to the model.

## 2.2.3. Bottleneck #3: Shared Mental Model Lock-In

Screening panels that evaluate ideas together over time converge on implicit criteria for what counts as "promising." These criteria are rarely made explicit, but are absorbed by new members through observation and experience of having ideas endorsed or rejected (DiMaggio & Powell, 1983; Mathieu et al., 2000). The result is a self-reinforcing evaluative culture. This constraint operates at the group level and can in principle compound over time, making it distinct from individual cognitive biases.

### *How Generative AI Interacts with the Bottlenecks*

Generative AI also exacerbates shared mental model lock-in, by changing both its substrate and its reach. The substrate shifts from a group of evaluators to a model: a screening AI trained on historical panel decisions learns and replicates the panel's implicit evaluative criteria at scale, institutionalizing those criteria without anyone having articulated or endorsed them. What was previously a constraint that could be disrupted by a dissenting panel member or a new hire with different priors becomes encoded in the system itself — and thus harder to detect, because the AI output looks like a principled judgment.

The reach extends from group deliberation moments to every individual decision the AI assists, depending on how this assistance is instantiated. A recent field experiment found that pairing AI pass/fail screening recommendations with a written rationale increased evaluators' compliance with rejection recommendations more than with acceptance recommendations, without improving their ability to discriminate between correct and incorrect AI judgments (Lane et al., 2026). The mechanism goes beyond over-reliance on AI: evaluators who were given AI rationales engaged less with the underlying submissions and were less likely to override AI recommendations even when doing so would have improved decision quality, because the rationale substituted for independent verification rather than supporting it. The effect was strongest for borderline cases, which are precisely where human judgment would be more valuable. Confident AI rationales thus furnish evaluators with ready-made justifications for dismissing the ideas the innovation process most needs to advance. The bottleneck is the

same—implicit, unarticulated convergence on what counts as "promising"—but it now operates at the level of individual decisions rather than panel meetings, and its locus shifts from human evaluators, who can be challenged or replaced, to a pervasive model whose outputs feel authoritative.

### *Intervention*

For *AI-mediated lock-in*, interventions can constrain how the AI system is trained and how its outputs are presented to evaluators. Models trained directly on historical panel decisions inherit those panels' biases by construction, because the only labels available are pass/fail judgments produced by the very evaluators the AI is meant to augment. Where post-launch outcomes are observable (sales, adoption, downstream success metrics), training on these outcomes rather than on screening decisions breaks the inheritance loop. Where outcomes are not observable—the typical case for ideas that were rejected and never built—the more conservative intervention is to use the AI for diversification rather than judgment. That is, AI can generate the strongest possible case *for* an idea before evaluators score it, counteracting the asymmetric salience of downside risk (Lane et al., 2022). Separately, the findings of Lane et al. (2026) suggest that the cleanest design choice is to withhold rationales altogether in screening contexts. Black-box recommendations in their field experiment improved decision quality relative to human-only evaluation, while the same recommendations paired with narrative justifications did not, despite inducing higher overall compliance. More research is needed on whether there are ways to introduce explanations while ensuring that screeners continue to critically evaluate the underlying submissions too.

## 2.3. Preference Measurement and Consumer Insight Stage

Following idea generation and idea screening, ideas are typically evolved into concepts with detailed features and specifications. Testing the concepts requires understanding consumer preferences and needs. Multiple qualitative and quantitative methods have been refined over the years to gain and quantify these insights. We focus on two bottlenecks in preference measurement: the *authenticity gap*, and the *articulability gap.*

### 2.3.1. Bottleneck #1: The Authenticity Gap

The *authenticity gap* is the difficulty of capturing what consumers actually prefer and feel, rather than what they articulate when asked in a decontextualized research setting. We deliberately use "authenticity" rather than "fidelity" or "accuracy". The latter terms imply a well-defined ground truth waiting to be reproduced. Yet in practice preferences are not stored, but are constructed in the act of elicitation, and are sensitive to context, framing, comparison sets, and current affective state (Lichtenstein & Slovic, 2006; Bettman et al., 1998). For example, preference for a candle is constructed from the scent, the jar design, color, packaging, surrounding context, and the comparison set. A research design that strips away context to achieve experimental control strips away the same features that determine whether an innovation will be adopted. Moreover, consumer perception is often of a unified gestalt rather than the sum of decomposable attributes (Burnap et al., 2023). The whole is registered before it is decomposed, and integration across modalities — visual, tactile, olfactory — is part of how the response is constructed. This pre-analytical, configural quality makes aesthetic experience particularly difficult for systems like LLMs, which learn primarily from language, to capture.

Finally, the authenticity gap also arises because consumer adoption of genuinely new products is disproportionately governed by nonnormative behaviors that are difficult to predict. These include the sunk cost fallacy that sustains loyalty to incumbent solutions even after they have failed; status quo bias, which makes the costs of switching feel larger than equivalent gains from the new product (Samuelson & Zeckhauser, 1988); and loss framing, whereby consumers weigh the certain losses involved in abandoning a familiar product more heavily than the uncertain benefits of the alternative (Kahneman & Tversky, 1979).

### *How Generative AI Interacts with the Bottleneck*

Generative AI interacts with the *authenticity gap* through three distinct mechanisms in LLM-based simulation of consumer responses.

The first is *heterogeneity loss*. This arises for example in one of the most direct applications of generative AI to preference measurement, LLM-based conjoint. Here the model is queried many times under different product configurations, prices, and (sometimes) demographic descriptors, and the resulting choice distributions are used to estimate willingness to pay, similarly to how one would do so with human conjoint data. Because LLMs are trained on population-level text data, they tend to reveal the central tendency of expressed preferences rather than the distribution of individual preferences within the population. The resulting estimates approximate average population preferences reasonably well but fail to capture the demographic and psychographic variation that drives segmentation, targeting, and positioning decisions. Even when fine-tuned on demographically stratified human conjoint data, the model recovers population averages but differences across groups are incoherent and do not reflect meaningful differences across segments (Brand et al., 2026). Authentic consumer research is compromised if the tool fails to capture heterogeneity relevant to whether people will adopt innovations.

The second is *prompt-induced confounding*, which arises whenever generative AI is used to simulate consumer responses to a manipulation, whether in aggregate conjoint or in more elaborate persona-based designs. Standard experimental practice requires that participants be blind to the manipulation–say, to variations in product price across conditions to estimate price sensitivity–but this convention creates a fundamental problem when applied to LLMs. Simulated subjects and contexts are generated dynamically from prompts, not drawn from a pre-existing population. When a variable like price is varied in a blind prompt, the model does not treat that variation as an exogenous manipulation; it treats it as informative of other contextual variables, just as they co-vary in real observational data. Gui and Toubia (2023) demonstrate this by benchmarking GPT-4 against an experiment in which human respondents evaluate consumer packaged goods at randomized prices. While the human data produce the expected downward-sloping demand curve, the LLM data produce implausibly flat or upward-sloping curves. The model reads higher prices as signaling a premium context rather than an experimental manipulation, yielding a directionally incorrect inference about the preference that the measure is meant to estimate.

The third is the *failure to reproduce non-normative behavior* in synthetic data, and in particular digital twins. A more ambitious approach to capturing authentic preferences than aggregate LLM-based conjoint is to construct simulations of specific individuals built from rich behavioral profiles of those people, including prior question responses, personality measures, and economic preferences (Toubia et al., 2025). Even with this much richer input, one implementation of such twins failed to reproduce many of the non-normative patterns that govern adoption of genuinely new products (Toubia et al., 2025; Peng et al., 2026). The digital twins choose normative options far more often than human participants, do not exhibit the sunk cost fallacy, and fail to violate the independence axiom of utility theory in the way real consumers systematically do. For a testing method to predict adoption dynamics, it needs to behave irrationally where consumers would.

*Intervention*

We address each of the complications in turn. Two admit partial fixes operating on the AI system; the other requires the appropriate intervention to remain on the human side.

For *heterogeneity loss*, no current intervention recovers meaningful between-group variation in LLM-based conjoint, but the bottleneck is less binding when aggregate-level estimates suffice. Brand et al. (2026) investigate the alignment of willingness-to-pay recovered from LLM-based conjoint with willingness-to-pay recovered from human studies. They find that for products and attributes that already exist, baseline LLM-based conjoint often produces willingness-to-pay estimates of the right order of magnitude at a fraction of the cost and time of a human study. For incremental new features within an established category—a projector added to a laptop, a new flavor in an existing line—fine-tuning on historical human conjoint data for that category (that excludes the new features) recovers much of the aggregate accuracy that the baseline misses. By fine-tuning on historical data, the LLM learns preferences in these data that then improve out-of-sample predictions because it has a better sense of the specific target population choices relative to the whole population it was initially trained on. The fix does not generalize across categories, however, and it does not recover meaningful heterogeneity even when the fine-tuning data is demographically stratified. Instead, it makes different groups more similar to each other after fine-tuning, reflecting the average across all populations instead of each group's preferences.

A complementary direction uses generative AI to augment rather than replace human surveys. These approaches treat LLM-generated responses as auxiliary data disciplined by a smaller human benchmark sample, rather than as a representative synthetic population. Ludwig et al., (2026) provide the econometric logic for this move: LLM-generated measurements can support estimation only when their errors are assessed against validation data. In conjoint settings,

Wang et al. (2026) show that combining human and LLM-generated responses through a debiased data-augmentation estimator can reduce estimation error and survey costs, while naive substitution can worsen bias. Brynjolfsson et al. (2026) similarly find that synthetic survey augmentation better approximates real consumer responses when LLMs receive rich contextual information from complementing individual responses, though performance varies across goods and demographic groups. Leng et al. (2026) develop a related calibration approach that reweights LLM responses across demographic personas to better match aggregate human benchmarks. These studies do not recover heterogeneity in the strong sense, but they identify a narrower and more defensible role for LLMs: supplementing small human samples with calibrated synthetic responses when the estimand is aggregate and the synthetic data are empirically calibrated.

For *prompt-induced confounding*, the intervention is on the design of the prompt itself, but it comes with a trap. The natural fix is to specify the contextual variables explicitly—tell the model the competing product's price, the store type, the competitive environment. But specifying these variables makes them artificially salient to the simulated consumer in a way they would not be in real decision-making. Returning to Gui and Toubia (2023): when the competing product's price is specified, the simulated demand curve collapses into a step function—purchase probability 1 below the competitor's price, 0 above it—missing the smooth downward slope observed in real consumers, most of whom are not attentive to competing prices on routine purchases. Two failure modes thus trade off: confounding when the manipulation is blinded, focalism when it is unblinded. The intervention that threads the needle is to unblind the *experimental design* rather than the contextual variables: tell the model that a variable is being experimentally varied across conditions, and specify the range of values it might take. Unblinding in this sense consistently improves simulation accuracy across model generations and is robust to irrelevant observational data in fine-tuning (Gui & Toubia, 2023). Combined with the previous mechanism, this points to a reasonable role for LLM-based simulation: a preliminary filter rather than a final verdict—winnowing a large candidate set down to a smaller one that merits serious human evaluation.

For the *failure to reproduce non-normative behavior* in digital twins, no current intervention yet restores the irrational patterns that govern adoption of novel products. Digital twins are not useless, but they should not be trusted on precisely the behaviors that distinguish adoption of genuinely new products from incremental ones. Decisions whose outcomes depend for example on the sunk cost fallacy, omission bias, or loss framing remain the province of human research, including the lead-user panels and ethnographic methods called for in Section 2.1.3.

### 2.3.1. Bottleneck #2: The Articulability Gap

The *articulability gap* concerns preferences consumers cannot articulate, often because they have never encountered a solution that would make the preference salient. We call these "hidden gems:" customer preferences that exist in practice but have not yet found a vocabulary because no product has yet demonstrated that the want is satisfiable. The canonical example is the touchscreen smartphone. Consumers in 2006 could not have articulated a preference for swipe-based navigation and touchscreens, because the possibility had not yet entered their experience. This is the demand-side analog of radical innovation. Just as radical product innovation requires escaping existing mental models of what a product can be, radical preference research requires identifying preferences that consumers themselves do not yet have language for. Standard qualitative and quantitative methods are structurally limited here: surveys and interviews can only surface what respondents can formulate in response to a prompt, and focus groups converge on preferences that participants can already express and compare. The methods used to uncover hidden gems are ethnographic and observational. They involve watching consumers work around the failure of existing solutions, before they have articulated what a better solution would look like.

### *How Generative AI Interacts with the Bottlenecks*

Generative AI exacerbates the *articulability gap* through a distinct route. AI systems trained on language can only surface preferences that have been expressed in language somewhere in the training corpus, so a preference that has never been articulated is by construction absent. Fine-tuning on reviews, interview transcripts, or support tickets expands the vocabulary of expressed preferences but does not reach the layer of experience that precedes expression—an architectural constraint that further scale will not resolve. A related concern arises with AI-moderated interviews and chatbots used to elicit preferences from real consumers: through the vocabulary they use, the order in which they probe, and the framing they impose, AI interviewers might consistently construct a different kind of preference than human interviewers do, e.g., more articulable, more context-stripped, more central-tendency in character. More research is needed to understand whether or not firms that shift to AI-moderated research at scale yield consumer insights that become systematically narrower than human-moderated research.

### *Intervention*

For the *articulability gap*, to the best of our knowledge the structural component admits no AI fix: ethnographic and observational methods that watch consumers work around the failure of existing solutions, remaining the only route to hidden gems. Generative AI does, however, contribute at the articulable layer through two supporting roles.

First, it can activate the researcher's own tacit understanding by serving as a probe rather than a measurement. For example, aesthetic and embodied preferences are difficult to decompose or describe in words; the response is felt before it is analyzed, even by experienced managers. When such managers treat generative AI outputs as stimuli, they can surface intuitions they carry but had not thought to articulate. In Burnap et al. (2023), machine-generated automotive designs shown to experienced design managers elicited explicit design ideas about specific dimensions, such as front overhang or hood slope—dimensions the managers had registered implicitly but had not put into words until a generated image brought them into relief. A related contribution comes from using generative AI as a simulated interviewee: when prompted to roleplay a consumer with specific characteristics and then interviewed by the researcher, the model can surface hypotheses the researcher had not previously formulated. This is because engaging with a plausible simulation (even if it is not a true simulation of a consumer) may activate the researcher's own hypothesizing in ways internal reflection alone does not (De Freitas et al., 2025).

Second, generative AI provides scalable access to expressed customer needs. Customer language in reviews, interview transcripts, and call center logs carries rich signals, but the raw language is specific, idiosyncratic, and often framed as complaints or feature requests rather than needs. Converting raw language into useful need statements (sometimes called *jobs to be done*) has historically required trained analysts. A consumer who says, "I couldn't tell which parts of the deck I'd already stained, so I kept missing spots," becomes a need statement like "able to see where I have already applied the stain." Models fine-tuned on professionally abstracted examples can perform this transformation at least as well as professional analysts, and the performance generalizes to new categories because the models learn the form of the transformation rather than the content of any domain (Timoshenko et al., 2026). This provides scalable access to the articulable layer while leaving the hidden gems to ethnography.

## 2.4. Diffusion and Market Learning Stage

Innovation does not succeed solely by virtue of being a good idea. Once developed, it must propagate through a social system to reach scale, and the signals that propagation generates must be read and converted into the next round of product decisions. These two activities—anticipating diffusion before and during launch, and learning from market signals afterward—define the diffusion and market-learning stage.

We discuss three bottlenecks. The first, *social diffusion modeling*, concerns anticipating how an innovation will propagate through a social system. The remaining two operate on the learning side: *attention and aggregation*, which concerns the volume of post-launch signal, and *motivated interpretation*, which concerns how organizations process that signal under incentive pressure to preserve prior decisions.

### 2.4.1. Bottleneck #1: Social Diffusion Modeling

Adoption is not the sum of independent individual responses. It is a social process in which early adopters generate the proof that makes the product credible to mainstream adopters, opinion leaders confer legitimacy on the category, and the product accumulates social-identity meanings as it moves through reference groups (Rogers, 2003). Anticipating this dynamic for a specific innovation in a specific market is what determines whether the firm should invest in scaling, repositioning, or abandoning a launch. The relevant unit of analysis is the social system rather than the consumer, which makes the work harder than modeling any individual response and which scales poorly: a launch team that can model the dynamics within one segment usually cannot model five, and the dynamics that matter most for a novel product often involve segments the firm has never directly studied.

### *How Generative AI Interacts with the Bottlenecks*

For *social diffusion modeling*, agent-based simulation of diffusion is the natural application, and recent pipelines for generating simulated agent populations (e.g., Park et al., 2023) have renewed interest in the long-standing literature on agent-based diffusion modeling (e.g., Goldenberg et al., 2002). The limits here, however, follow directly from Section 2.3: if individual consumer behavior cannot yet be simulated authentically—particularly the non-normative behaviors that govern adoption of genuinely novel products—then social diffusion built from those individuals inherits the same problem, and likely amplifies it as errors propagate through interaction effects. We do not belabor this point; it is the same bottleneck appearing at a higher level of aggregation.

### *Intervention*

The interventions in this stage are less developed than for the earlier stages, in part because the research base is thinner. For *social diffusion modeling*, to the best of our knowledge no AI-side intervention currently resolves the bottleneck, and the constraint follows directly from Section 2.3: simulations cannot reproduce the non-normative behaviors that govern adoption of genuinely novel products. The appropriate role for AI is the same one identified for preference measurement—screening plausible adoption paths rather than adjudicating which will materialize—with the consequential calls remaining on the human side, supported by lead-user and ethnographic work.

## 2.4.1. Bottleneck #2: Attention and Aggregation

Once a product is in the market, it emits signals at volumes no human team can easily absorb—reviews across multiple platforms, social media discussion, support tickets, returns, community forums, search queries, and sales by segment. Unstructured consumer text has long been recognized as a rich source of market-structure insights and customer needs, with findings comparable in quality to traditional interview-based research (Netzer et al., 2012; Timoshenko & Hauser, 2019), but extracting that signal has historically required specialist teams and substantial processing time, making the data effectively inaccessible for rapid innovation cycles. The volume problem is compounded by selection bias in what surfaces: reviews overrepresent extremes of satisfaction and dissatisfaction (Schoenmueller et al., 2020), early reviewers hold systematically different preferences than later buyers (Li & Hitt, 2008), social media overrepresents the engaged and the articulate, and support tickets overrepresent failure modes. The signals that matter most for innovation—emerging use cases or silent non-adoption—are precisely the ones least likely to surface.

### *How Generative AI Interacts with the Bottleneck*

For *attention and aggregation*, generative AI dramatically alleviates the volume constraint. LLMs can ingest, classify, cluster, and summarize market feedback at a scale previously feasible only for the largest firms with dedicated insights teams, and models can extract structured customer needs with quality approaching professional analysts (Timoshenko et al., 2026). The cost and latency of market listening have collapsed. The bottleneck is not fully resolved but shifted, however. Product teams now face a prioritization problem: which of the hundreds of themes surfaced this week reflect signals worth acting on, which are artifacts of the summarization process, and which are weak signals of something important that a frequency-weighted summary will systematically hide?

### *Intervention*

For *attention and aggregation*, generative AI already provides substantial relief through scalable extraction of structured needs from unstructured text (Timoshenko et al., 2026). The harder open question is how to architect triage so that what surfaces is the signal that matters for the next round of innovation rather than the signal that is most frequent. Adjacent work on LLM-assisted screening of crowdsourced submissions suggests that combining model ratings with historical expert judgment can preserve the ability to surface promising outliers (Kireyev et al., 2025), but the transfer of this architecture from pre-launch screening to post-launch review has not been tested directly. Whether it can be tuned to surface weak signals about emerging use cases or silent non-adoption—rather than the loudest themes in the corpus—is an open question.

### 2.4.1. Bottleneck #3: Motivated Interpretation

Even with the signal in hand, organizations have incentives to read it charitably toward decisions already made, consistent with motivated-reasoning accounts of evidence evaluation (Kunda, 1990). Weak negative signals are often assimilated into narratives that preserve the incumbent product logic, while positive signals are disproportionately amplified: managers may overuse initial positive beliefs when interpreting later negative information and persist with failing new products despite improved information about their prospects (Boulding et al., 1997; Biyalogorsky et al., 2006; Liang, 2021). This behavior stems from sunk investments and high-stakes organizational politics rather than a lack of expertise. Providing better data does not resolve the constraint if the underlying incentives to charitably interpret that data persist.

#### *How Generative AI Interacts with the Bottleneck*

For *motivated interpretation*, generative AI has a peculiar two-sided relationship with the bottleneck. On one side, an AI system has no ego investment in past decisions and will surface a pattern of complaints about a championed feature with the same priority as a pattern of praise. On the other, the fluent synthesis it produces is easier to selectively quote in support of conclusions the team was already inclined toward. The authority of an AI-generated summary may make motivated reasoning easier to dress up rather than harder to sustain.

*Intervention*

For *motivated interpretation*, we know of no direct empirical work on whether AI-mediated synthesis of post-launch market signal amplifies or counteracts confirmation bias in organizational decision-making. As discussed just above, AI has no ego investment in past decisions but its fluent output is also easier to selectively cite. Whether these forces net out toward more honest interpretation or more sophisticated rationalization is, we suspect, the most important open empirical question in this part of the framework.

## 2.5. Discussion

The preceding analysis points to a division of labor between humans and AI organized not by stage but by the mechanism behind each bottleneck. Where the constraint is informational, volumetric, or methodological—cognitive fixation in ideation, cognitive load in screening, attention and aggregation in market learning—the appropriate intervention is on the AI system itself, through prompting strategies, fine-tuning, or information architecture. Where the constraint is in tacit lived experience, non-normative behavior, or motivated interpretation—design fixation across an industry, the articulability gap, the failure of digital twins to reproduce irrational adoption dynamics, the limits of social diffusion modeling, and the organizational incentives that distort post-launch signal—the appropriate intervention remains (for now) on the human side, with generative AI in a supporting role at best.

This division is not a permanent map. It is what the mechanisms imply given current capabilities. The framework's diagnostic value lies in this conditionality: naming the mechanism behind a bottleneck lets a manager or researcher predict whether a new AI capability is likely to alleviate the constraint. The mechanism, not the capability, determines the direction of the effect.

## 3. Broader Systemic Implications of Gen AI on Innovation

Section 2 analyzed how generative AI interacts with specific human bottlenecks at each stage of the innovation process. This section zooms out to consider three broader issues that are relevant to the innovation system as a whole rather than any single stage: *organizational issues*, *deskilling and automation trap*, and *inequality and representation*. Finally, we discuss three ways in which GenAI may call for a re-thinking of the innovation process: by inverting the development curve, by changing the architecture of the innovation process, and by forcing us to consider innovation *for* AI agents. The three issues discussed in Section 3.1 as well as the three potential disruptions discussed in Section 3.2 each present a paradigm shift in innovation with its own associated challenges. Table 2 summarizes each subsection in terms of the paradigm shift it identifies, the challenge it poses, and the proposed direction for practitioners and researchers.

**Table 2.** Systemic implications of generative AI on innovation.

| Topic | Paradigm Shift | Challenge | Proposed Direction |
| --- | --- | --- | --- |
| Organizational Issues | Technical barrier to building AI-enabled tools has fallen | Handing development of AI tools entirely to functional experts leads to unsustainable tools | Couple centralized technical infrastructure with distributed collaborative building |
| Deskilling & Automation Trap | Automation of adjacent innovation tasks | Erosion of the human capabilities the framework relies on | Maintain meaningful contact with human experience at each stage |
| Inequality & Representation | AI optimizes for the measurable and the majority | Heterogeneity loss compounds into market-level under-representation | Identify where AI adoption deepens versus alleviates representation gaps |
| Inverted Development Curve | Prototyping is cheap; reliable deployment is hard | Stage-gate screening fails because the real de-risking happens post-prototype | Move evaluation infrastructure to the post-prototype stage |
| The Changing Architecture of Innovation | AI agents primarily treated as integrated innovators | Loss of authentic needs information | Leverage AI for needs elicitation and as user innovation toolkits |
| Innovating for AI Agents | AI agents are becoming end users of innovation | Innovation methods built for humans may not apply | Rethink customer insight and preference measurement for AI agents |

### 3.1. Broader Implications of Generative AI for the Innovation Process

#### 3.1.1. Organizational Issues

The bottleneck framework in Section 2 treats generative AI as a set of interventions applied to specific points in the innovation process. However, whether those interventions are available to a firm at all depends on a prior organizational question: who inside the firm is permitted to build, shape, and deploy AI tools, and how does that capability travel across functions? The interventions that matter most are not off-the-shelf products. Technologies such as persona-prompted ideation sessions, unblinded simulation designs, and fine-tuned need-extraction pipelines are custom configurations that require domain expertise. Consumer insights teams, product managers, and designers hold this expertise but data science teams typically do not. In contrast, data science teams often own the technical infrastructure but consumer insights teams are typically less familiar with it.

What makes this moment different is that the technical barrier to building AI-enabled tools has fallen dramatically. Natural-language interfaces and coding assistants now allow people without formal engineering training to build working proof-of-concepts. The functional experts closest to the innovation problem, such as consumer insights researchers, are increasingly able to participate directly in producing the logic of the tools they use, consistent with evidence that AI and algorithms create greater value when algorithmic literacy is broadly dispersed among domain experts rather than concentrated inside IT (Tambe, 2026). However, this widespread ability to generate code creates a dangerous illusion that anyone can deploy enterprise software. Machine-learning systems are especially prone to hidden technical debt: early prototypes can be easy to build but costly to maintain once they become embedded in production systems (Sculley et al., 2015). Domain experts possess deep understanding of the applications but lack the software product management experience required to oversee the lifecycle, user interface, and maintenance of a deployed tool. While this shift erodes the traditional justification for keeping AI capability inside a dedicated technical silo, it does not eliminate the need for engineering rigor. Firms that retain the old architecture can actively foreclose new modes of value creation, yet handing deployment entirely to functional experts leads to unsustainable tools. One possible solution couples a centralized technical infrastructure with distributed collaborative building. Consumer insights experts co-create the logic and initial prototypes, data scientists refine the underlying models, and dedicated software engineers manage the deployment, governance, and substrate (Iansiti & Lakhani, 2020).

Relatedly, competitive advantage in AI-enabled innovation comes less from the sophistication of any specific model. Models are increasingly commoditized, since the same foundation models are available to most entrants (De Freitas, 2023). Instead, advantage comes from having unique data and two integration capabilities. The first is a federated approach in which those with domain knowledge are involved in AI tool configuration decisions: choices about fine-tuning data, prompts, guardrails, and what counts as good output are technical to implement but substantive to get right. The people who know what a good output looks like sit in insights, product, and design, not in data science or IT. The second is deploying the resulting tools deeply enough into workflows that they change what gets done. This is consistent with the broader IT productivity literature, which shows that value from digital technologies depends heavily on complementary organizational investments and changes in work practices, not just the technology itself (Brynjolfsson & Hitt, 2000). The first requires functional experts to co-own configuration rather than hand off requirements, and the second requires rewiring decision processes, not just adding tools to them.

The relevant organizational question concerns decision rights and accountability: who has the authority to configure AI systems, interpret their outputs, and own the resulting decisions (Sambamurthy and Zmud, 1999). Goldberg and Puranam (2026) argue that AI displacement depends not only on technical performance but also on the normative appropriateness of delegating a task to AI. The same logic applies inside the innovation process. Even when AI systems can generate outputs that are technically useful, firms still need to decide which actors have the authority to configure those systems, interpret their outputs, and own the resulting calls. The organizational design problem is therefore not to automate the innovation function, but to allocate AI configuration, domain judgment, and accountability across functions.

Further, recall from the framework of Section 2 the most valuable human contributions to innovation: lead user identification, ethnographic observation, interpretation of weak post-launch signals, and accountability for consequential calls under uncertainty. This suggests that the consumer insights team should act as a co-producer of AI tooling calibrated to those capabilities, rather than merely a recipient of AI outputs generated elsewhere. Insights teams that remain on the consumer side of the silo will find that tooling built without their involvement encodes assumptions about consumer behavior that the function itself would have flagged as incorrect, for example, models of how customers respond to price changes or which segments share which preferences.

The organizational challenge is to both break down the silo between insights and data science and do so in a way that preserves the domain authority of the insights function rather than subordinating it to technical priorities.

### 3.1.2. Deskilling and Automation Trap

As generative AI takes over adjacent tasks in the innovation process, the human capabilities those tasks used to develop may quietly erode, aka *deskilling*. This has a compounding generational dimension: if generative AI generates ideas, simulates consumers, and drafts communications, junior practitioners lose the formative work that develops tacit judgment. The organization of the future may contain sophisticated users of AI-generated outputs who lack the judgment to recognize when those outputs are wrong. Unlike ordinary deskilling, this capability gap remains invisible until a genuinely novel problem requires grounded judgment that no one in the organization has had to build.

The bottlenecks reviewed in Section 2 all implicate the same accountability question: who owns the innovation? Genuine novelty often requires someone who has lived the problem. Honest preference measurement requires someone whose credibility with the consumer is real. This distinction mirrors Goldberg and Puranam’s (2026) argument that AI may become increasingly capable at prediction and coordination, but managerial work remains necessary where the task is to define goals worth pursuing and standing behind. The implied division of labor is consistent with the framework: generative AI should be targeted at informational, volumetric, and methodological tasks, while humans must be deliberately protected in roles involving lead-user insight, radical reframing, and accountability for consequences.

These losses compound into a systemic risk, aka *the automation trap*. As AI tools take on ideation, screening, preference simulation, and consumer interviews simultaneously, the feedback signals organizations use to learn what works become increasingly AI-produced. Individual tools may be accurate in isolation, but the learning loop gradually loses contact with actual consumer behavior, producing generic outputs that lack real sensibility about real consumers (Huang & Rust, 2025). Teams can iterate rapidly through AI-generated feedback cycles—refining concepts against simulated reactions, screening with AI-trained criteria, testing on digital twins—while the system as a whole drifts from the ground truth it was designed to track. Firms must preserve touchpoints with real consumers, real markets, and real signals at each stage, rather than assume that accurate components add up to a grounded system.

#### 3.1.3. Inequality and Representation

Discussions of generative AI often focus on uneven access to the technology. A more consequential inequality for innovation operates on the output side, as a downstream consequence of the heterogeneity-loss bottleneck identified in Section 2.3. Because AI systems are trained predominantly on data from majority populations, they implicitly optimize for those groups. Digital twins, for instance, are more accurate for affluent, highly educated consumers (Peng et al., 2026), and LLM-based conjoint recovers population averages but not between-group differences even after demographically stratified fine-tuning (Brand et al., 2026). Much like popularity bias in recommender systems (Carnovalini et al., 2025), LLMs approximate average preferences while erasing the variation that drives niche segmentation. At the level of an innovation system that uses these methods to allocate research effort across segments, it compounds into market-level under-representation: the segments whose preferences are hardest for current AI tools to reproduce are the segments whose unmet needs are least likely to surface as targets for new product development. Even when platforms try to correct for these biases, they may disproportionately focus on politically- salient groups (e.g., based on race or gender in the U.S.) rather than various other minority populations (Samure et al., 2025).

The interventions in Section 2.3.3—augmenting rather than replacing human surveys, fine-tuning on category-specific human data, using AI-moderated interviews to extend qualitative reach—are available levers. How to deploy them across the innovation portfolio to protect against representation collapse is an open empirical question.

### 3.2. Re-thinking the Innovation Process in the Age of Gen AI

Beyond modifying individual bottlenecks, generative AI puts pressure on the structure of the innovation process itself. This section considers three ways in which the traditional sequence of stages may need to be rethought: the inversion of development costs, a shift in the architecture of who participates in innovation, and the emergence of AI agents as end users.

#### 3.2.1. The Inverted Development Curve

In traditional software or product development, reaching a working prototype was slow and costly, but once it existed, the path to a reliable product was comparatively well-understood: inputs mapped to outputs according to rules the developer controlled. Products built around generative AI invert this curve. Reaching a prototype is now fast, as an LLM or LLM-built product will respond plausibly to most queries almost immediately, but moving from a plausible response to a reliable deployment is substantially harder–Figure 1. Failure modes are difficult to enumerate in advance, the same prompt can yield different outputs across runs, and domain-specific accuracy requires architectural choices about retrieval, grounding, validation, and fallback behavior that constitute most of the engineering effort.

The screening bottleneck therefore shifts in location. Section 2.2 considered stage-gate panels evaluating early-stage ideas, where the question is whether to advance an idea at all. Here the question is different: given a working prototype, can the system be made reliable enough to deploy? This judgment cannot be made by inspecting the prototype, since the demo is no longer informative about deployment risk. It requires evaluating organizational capacity for the post-prototype work—domain data assembly, validation architecture, integration—that constitutes most of the engineering effort but is hidden from the panels of judges currently in place to make this call.

Three consequences follow. First, firms accumulate stalled pilots: financial frameworks calibrated to the traditional curve fund a growing portfolio of pilots that never graduate. Second, stage-gate processes lose informational value, because when the hardest engineering happens after the prototype, an early gate carries little information about deployment risk. Third, competitive advantage shifts. Prototyping is commoditized because the same foundation models are available to most entrants, and differentiation now lives in the post-prototype work. Familiar triage heuristics such as demo quality and prototyping speed systematically misjudge value. The more informative signal is whether a firm has the post-prototype infrastructure to move a system from plausible to reliable in a specific domain.

**Figure 1.** The inverted development curve.

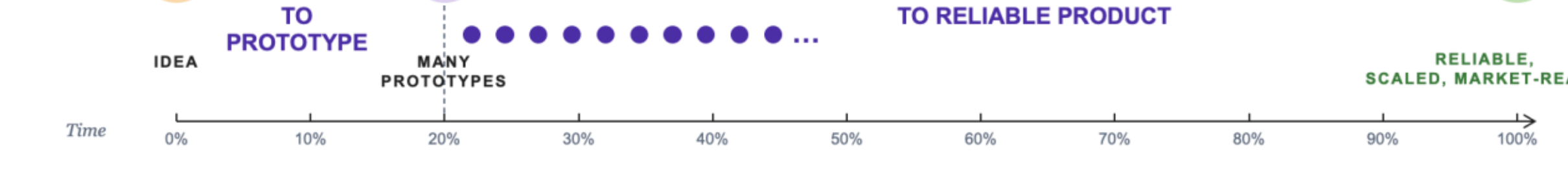

### 3.2.2. The Changing Architecture of Innovation

This subsection considers how AI alters the architecture of the innovation process itself. We apply Von Hippel's (1994) "sticky information" framework—which posits that information about user needs and product solutions typically resides in different places and is hard to transfer— to distinguish three distinct modes in which generative AI can be deployed. Most current attention has focused on the first, but the durable value of generative AI in innovation may lie in the other two.

*Mode 1: AI as integrated innovator.* Many generative AI solutions attempt to make the AI system itself the locus of both needs and solutions information, drawing on training data, in-context learning, retrieval-augmented generation, or fine-tuning. A synthetic persona designed to capture the preferences of a customer segment is one example. Section 2 highlighted the limitations of this approach: heterogeneity loss in preference simulation, the failure of digital twins to reproduce non-normative behavior, and the deeper authenticity bottleneck that arises because consumer preferences are constructed rather than stored. These limitations suggest the locus of needs information should, in many cases, remain with actual users.

*Mode 2: AI-moderated needs elicitation.* Rather than substituting for users, generative AI can extract information about their needs at scale. As mentioned earlier, AI-moderated interviews use AI agents to interview human consumers, potentially uncovering rich insights with reduced budget and time (Korst et al., 2026). The approach is receiving considerable industry attention but remains relatively untested in academia. Research is needed on its authenticity, accuracy, and reliability, and on the conditions under which it may be trusted.

*Mode 3: AI as user innovation toolkit.* A third mode of human–AI collaboration makes solutions information accessible to users, empowering them to become innovators themselves (Peres et al., 2023). Many "vibe coding" or "low code" tools fit this pattern. The approach itself is not new—it was foreshadowed by Von Hippel's (2001) work on user toolkits—but generative AI extends it substantially. When users develop their own solutions, the role of firms becomes to curate and distribute those with mass-market potential, becoming "publishers" of user innovations (Von Hippel, 2006). Whether firms are still needed as publishers in the age of generative AI, given that barriers to deployment are also lowered, is an open question with welfare consequences.

Taken together, these modes (Figure 2) suggest the focus on Mode 1 may be disproportionate to its likely long-term value. The limitations established in Section 2 constrain what AI-as-integrated-innovator can reliably achieve, while Modes 2 and 3 work with rather than around the human constraints that make innovation hard. Identifying which mode fits which context, and developing tools tailored to each, is a priority for future work.

**Figure 2.** The changing architecture of innovation.

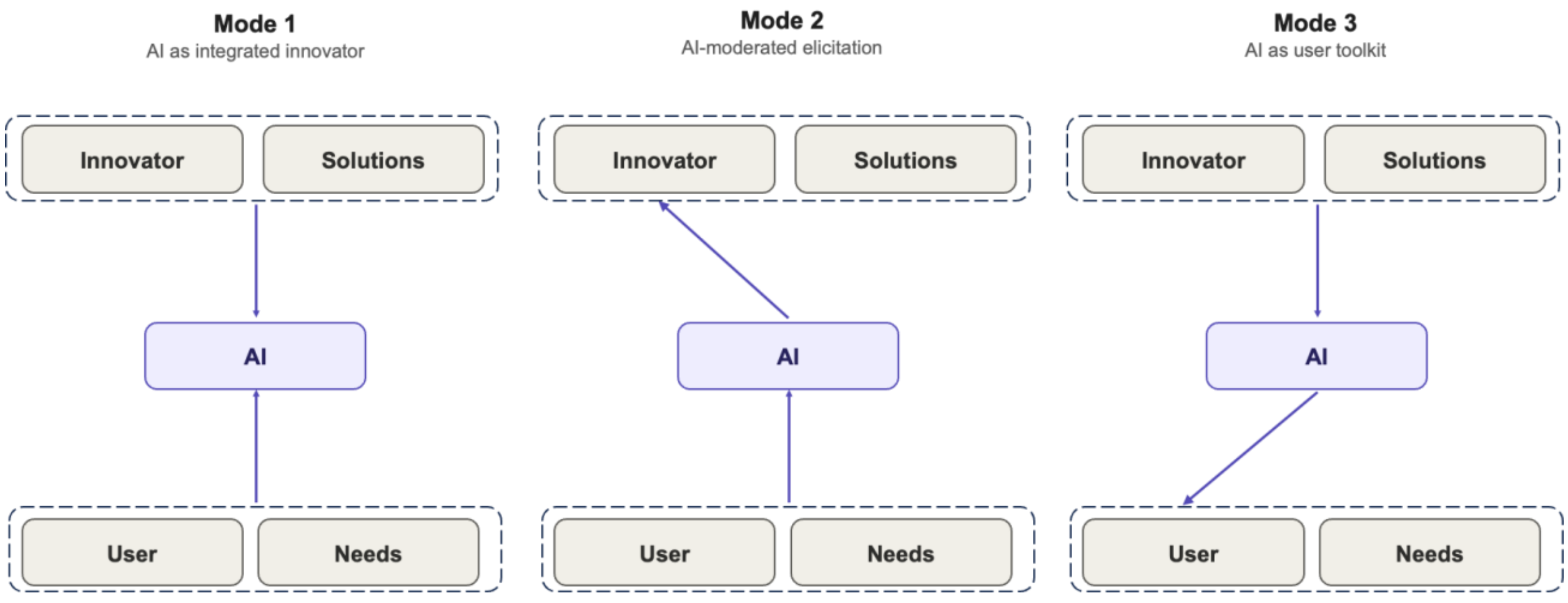

### 3.2.3. Innovating for AI Agents

Existing research on generative AI in innovation has focused on automating the innovation process while assuming the user is human. This subsection considers an only slightly futuristic scenario in which the user is an AI agent and the innovator is either a human or another AI agent.

Li et al. (2026) describe the internet as shifting from a human-centric content economy to a machine-centric agentic economy. Matthias Biilmann, CEO of Netlify, coined the term "Agent Experience (AX)" to describe the holistic experience AI agents have when interacting with products, platforms, or systems (Besson, 2025), and Wong et al. (2025) developed a software engineering platform that jointly optimizes for Agent Experience, User Experience, and Developer Experience, arguing that passing human-oriented content directly into an agent's prompts degrades agent experience.

This opens research opportunities for business scholars to redefine innovation when the customer is an AI agent. Should "Agent Insight" emerge as a discipline mirroring customer insights? How should preference measurement be adapted to AI agents? Methods developed for humans may not transfer; entirely new tools must be developed that are tailored to how AI agents are built (Gui & Toubia, 2023).

AI agents are also beginning to serve as both innovators and users. Kang and Yoganarasimhan (2025) develop a self-improving system that combines AI agents generating solutions with AI agents providing feedback, allowing the iterative feedback loop of innovation to become nearly instantaneous. Baptista and Nunes (2026) raise the question of how trust should be defined and measured in these contexts. AI innovators and users will need to remain embedded in a broader ecosystem that includes human oversight.

A more fundamental question is what "preference" means when the buyer is an AI agent. Human preferences are constructed, affective, and context-dependent. AI agent "preferences" are objective functions set by the humans or organizations that configure them. What looks like an AI agent's preference is a reflection of its configuration. This creates a principal-agent problem: the agent optimizes for the human's stated preferences as encoded in its configuration, not necessarily for the preferences the human would construct in context or the utility they would derive from the product in use. Optimizing for AI agent criteria may therefore diverge systematically from optimizing for human satisfaction, with the gap largest for the novel, high-involvement products that innovation is most concerned with.

## 4. Conclusions

The dominant question in discussions of generative AI and innovation has been *whether* humans will remain in the process. We have tried to reframe it as *when* and *how*. By grounding the analysis in human bottlenecks rather than AI capabilities, the framework identifies the specific points in the innovation process where human judgment is the binding constraint and the specific points where it is not. This distinction is empirical, following from the mechanism behind each bottleneck. Where the constraint is informational, volumetric, or methodological, generative AI is the appropriate intervention, and the gains come from designing the AI system well. Where the constraint is in tacit lived experience, non-normative behavior, or motivated interpretation, the constraint sits with humans and no foreseeable AI capability dissolves it. Naming the mechanism behind a bottleneck lets a manager or researcher predict, for any new development, which kind of constraint is in play and therefore which side of the division the development falls on.

The systemic implications of this division are harder to act on than the stage-level findings. The human contributions the framework relies on — lead-user immersion, ethnographic observation, willingness to challenge organizational narrative — are exactly the contributions that current organizational practice is least equipped to protect. They are developed through formative work that AI is rapidly absorbing, valued through accountability structures that AI diffuses, and exercised by people whose credibility with consumers AI cannot supply. Firms have largely settled the question of whether to use generative AI in the innovation process. The harder question is whether the human capabilities that process depends on will still be there when novel problems arise. Those capabilities are load-bearing, they erode under conditions that look like progress, and they require deliberate organizational protection.